\documentclass[conference]{IEEEtran}
\IEEEoverridecommandlockouts

\usepackage{cite}
\usepackage{amsmath,amssymb,amsfonts}
\usepackage{algorithmic}
\usepackage{graphicx}
\usepackage{textcomp}
\usepackage{xcolor}
\def\BibTeX{{\rm B\kern-.05em{\sc i\kern-.025em b}\kern-.08em
    T\kern-.1667em\lower.7ex\hbox{E}\kern-.125emX}}

\usepackage{lipsum}
\usepackage{float}

\usepackage{soul}
\usepackage{xcolor}
\usepackage{xspace}

\usepackage{hyperref}

\hypersetup{
    colorlinks=true,
    linkcolor=blue,
    urlcolor=violet,
    citecolor=black,
}

\begin{document}

\title{Evolving the Computational Notebook: \\ A Two-Dimensional Canvas for Enhanced Human-AI Interaction 
}

\author{
\IEEEauthorblockN{Konstantin Grotov}
\IEEEauthorblockA{\textit{Neapolis University Paphos}\\
Paphos, Cyprus}
\IEEEauthorblockA{
konstantin.grotov@gmail.com}
\and
\IEEEauthorblockN{Dmitry Botov}
\IEEEauthorblockA{\textit{Neapolis University Paphos}\\
Paphos, Cyprus}
\IEEEauthorblockA{\textit{AI Future Hub}\\
dmbotov@gmail.com}
}

\maketitle

\begin{abstract}
Computational notebooks, while essential for data science, are limited by their one-dimensional interface, which poorly aligns with non-linear developer workflows and complicates collaboration and human-AI interaction. In this work, we focus on features of Computational Canvas, a novel two-dimensional interface that evolves notebooks to enhance data analysis and AI-assisted development within integrated development environments (IDEs). We present vital features, including freely arrangeable code cells, separate environments, and improved output management. These features are designed to facilitate intuitive organization, visual exploration, and natural collaboration with other users and AI agents. We also show the implementation of Computational Canvas with designed features as a Visual Studio Code plugin. By shifting from linear to two-dimensional spatial interfaces, we aim to significantly boost developers' productivity in data exploration, experimentation, and AI-assisted development, addressing the current limitations of traditional notebooks and fostering more flexible, collaborative data science workflows. 

\end{abstract}


\section{Introduction}
\begin{figure*}
\centering
\includegraphics[width=1\textwidth]{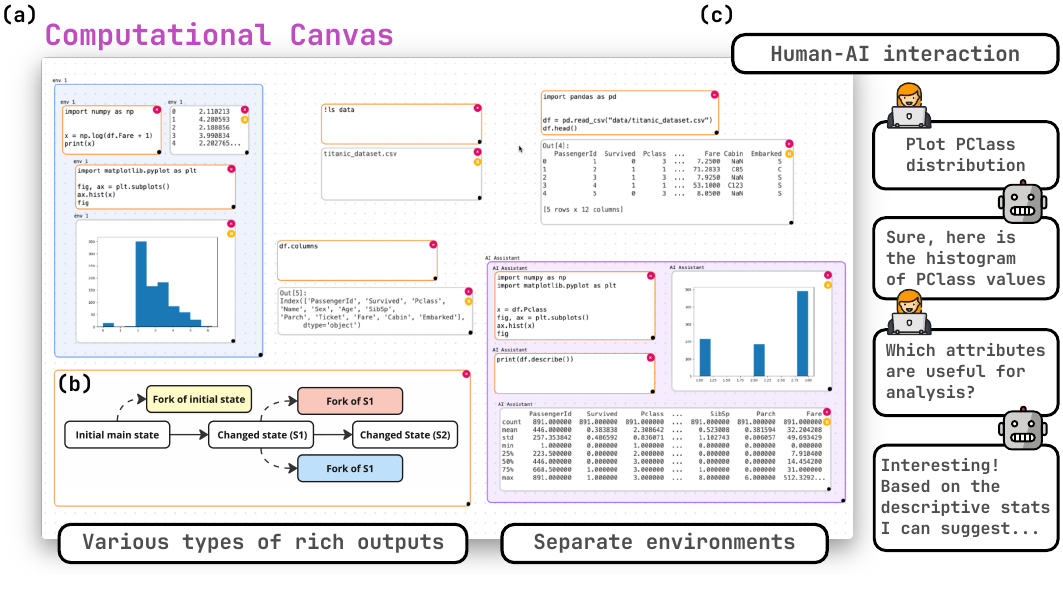}
\caption{\textbf{(a)} Prototype interface of the computational canvas. Orange cells in the canvas are the executable cells, the grey cells are the outputs, and the colorful areas are the separate environments with independent runtime states. \textbf{(b)} Schematic representation of the forking mechanism while creating a new environment. \textbf{(c)} A schematic representation of a chat between a user and an AI assistant is in Computational Canvas. The agent creates the dedicated environment that executes the user request.}
\label{fig}
\end{figure*}

Computational notebooks are the favored medium for writing code and follow the paradigm of literal programming~\cite{knuth1984literate}. Nowadays, computational notebooks have played a crucial role in different areas, like data analysis~\cite{perkel2018jupyter}, science~\cite{beg2021using, randles2017using}, and education~\cite{johnson2020benefits, barba2019teaching}. 

The framework Jupyter~\cite{Kluyver2016jupyter}, renowned for its computational notebooks and extensively utilized by data scientists~\cite{perkel2018jupyter}, is distinguished by its ability to develop code in a stateful manner. 
This means that each new code execution incrementally updates the notebook's runtime state, allowing users to iterate more swiftly without re-running all the previous code. 

Although computational notebooks are a convenient medium for data analysis, the code users write in them is often of worse quality and more entangled structure rather than the one they write in scripts~\cite{grotov2022large}. This entanglement saves beyond the single cell level, as the notebook cell order often doesn't correlate with the actual highly non-linear developer's workflow~\cite{chattopadhyay2020s, ramasamy2023workflow}. Moreover, notebooks struggle with very poor reproducibility~\cite{pimentel2019large, pimentel2021understanding}, especially in domains related to science~\cite{samuel2024computational}, where reproducibility is crucial. The problems mentioned above make notebooks a medium that is not reliable for end-to-end software engineering but is the perfect choice for retrieving insights from data and experimentation~\cite{ramasamy2023workflow, rule2018exploration}.

With the rise of artificial intelligence and foundational models~\cite{OpenAI_GPT4_2023, reid2024gemini}, the literate programming paradigm has been changing accordingly~\cite{shi2024natural}, making the initial notebook interface limited in providing a comprehensible human-AI interaction. AI assistants often generate code snippets, modify multiple interdependent cells, or suggest alternative approaches that require users to revisit and restructure existing code. In a linear notebook, these interactions can quickly become chaotic, leading to confusion about the execution order, unintended overwrites, and cognitive overload for users trying to keep track of changes~\cite{mcnutt2023design}.

Recent studies proposed developing notebooks to streamline the experimentation workflow and collaboration rather than transferring and mimicking the existing tools from integrated development environments (IDEs) to notebooks~\cite{titov2024hidden, wang2024don, zheng2022telling}. Moreover, the studies on changing the one-dimensional structure of the notebooks to a two-dimensional one showed a boost in developer productivity~\cite{christman20232d, harden2023there, harden2022exploring}. Although the mentioned studies highlighted user interest in two-dimensional notebooks, their implementation was limited to multicolumn layouts, so they did not fully explore the new functionalities that true two-dimensional notebooks could offer.

To fill this gap, in this work, we propose a specific set of features that could facilitate the developer's experimentation in Computational Canvas\footnote{\url{https://github.com/konstantgr/computational-canvas}} -- the new interface enhances the traditional computational notebook by providing a two-dimensional workspace. Our approach allows users to create multiple, safely separate environments that branch off from the main runtime state, making it easier to experiment within a unified space. In this work, we outline the key features of this interface and explain how it can facilitate human-AI interaction. Moreover, we have implemented the computational canvas with the designed features as a plugin for Visual Studio Code. This approach simplifies the system, fostering non-linear experimentation and seamless collaboration without conflicts right in the IDE. These collaborations can occur not only between developers but also with AI assistants.

\section{Computational Canvas}

The Computational Canvas is a platform that allows users to experiment with code and analyze data in a two-dimensional plane. The canvas prototype is elucidated in Figure~\ref{fig}(a). The canvas's core features include a flexible arrangement of executable cells, support for different rich output formats, and the creation of a separate environment with forks of the main runtime state. This section will describe these features in more detail.

\subsection{Cells Arrangement} 
Development in computational canvas is a process of creating executable cells with the code and obtaining the execution output, similar to workflow in the Jupyter Notebook. However, contrary to the vanilla Jupyter Notebook, the cells in the computational canvas are not bound to any grid. Any cell could be moved to the desired position of the developer, making it possible to organize the cells in groups more naturally than in a linear notebook. The interface was inspired by the Miro board\footnote{https://miro.com} interface, which is widely used for collaborative brainstorming.

It's worth mentioning that the computational canvas is designed to be infinite, so the user can separate their workflow and comfortably utilize the canvas by dividing it into the needed groups. In contrast with the common one-dimensional computational notebook, the interface does not implicitly bind the user to follow the linear workflow, thus facilitating non-linear development.

\subsection{Rich Outputs Manipulation}
After the cell is executed, an additional output cell is created below it. Like the standard notebook, the computational canvas supports textual and graphical outputs, making it possible to use it for data analysis, where visual representation is often crucial. Whenever the same code cell is executed again, the output cell will update accordingly, even if it is located in different positions on the canvas. This feature allows users to separate the canvas into the area of executable code cells and the area where the users expect to review the output, making the iterations with the code or debugging more comprehensive. 

\begin{figure*}
\centering
\includegraphics[width=0.95\textwidth]{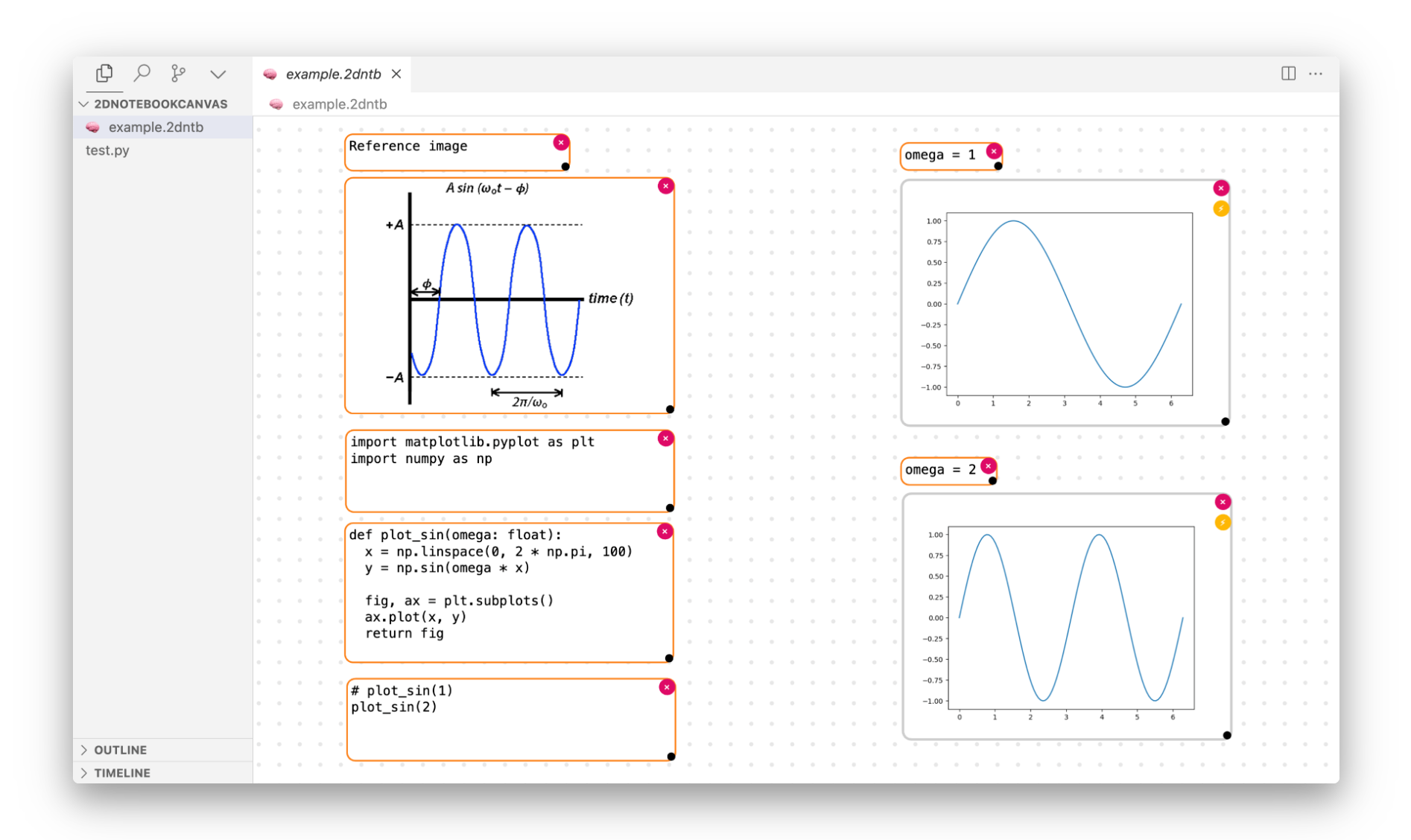}
\caption{example of detached outputs in the computational canvas opened in Visual Studio Code. The left side showed the code, which was executed with different parameters. The right side showed the outputs, which were detached and grouped together.}
\label{fig:detached-outputs}
\end{figure*}

Moreover, compared to the vanilla Jupyter Notebook, we're proposing the feature of detaching the output from its parent cell and fixing its state at the moment. On the one hand, it could help users develop and keep all gained insights in one place without worrying about occasionally rewriting them. Additionally, it provides a more structured way to organize outputs, which can be beneficial for presentations. Figure~\ref{fig:detached-outputs} illustrates an example of detached outputs.

\subsection{Separate Environments} 

The developer's workflow in computational notebooks is highly non-linear. By leveraging a two-dimensional space, users can break free from linear constraints. However, this brings another issue: ensuring that the code remains aligned with its intended purpose during the workflow.

For example, while working with a Pandas table \texttt{df}, numerous transformations and manipulations might be applied throughout the development process. As a result, code that was written and executed at the start may become obsolete or lead to unexpected results as the table evolves. Even worse, during the experimentation, crucial data could be corrupted, and the whole reproduction of the pipeline would be needed.

To address this issue, we propose a feature of separate environments within the computational canvas. This approach will enable developers to experiment freely within these environments while ensuring their main pipeline remains well-organized and traceable. 

Initially, the first cell execution in the canvas creates a main runtime – a process that stores the variables and functions declared in the previous executions. At any time, the user can create a new distinct environment, which will initially be forked from the main runtime but will not be referenced anymore. Then, this environment could be used for safe exploration and experimentation. Figure~\ref{fig}(b) shows the schematic forking diagram.

The user interface displays the environment as a colored, interactive area that can be moved around the canvas or resized. When placed in a particular environment, a new cell will be executed using its associated runtime. Since each environment is a distinct instance, it can be moved along with all the cells it contains. Figure~\ref{fig}(a) illustrates this with two environments, represented by purple and blue areas. 

\subsection{Collaborative Exploration}

The computational canvas could be a promising tool for more natural and collaborative exploration than the computational notebooks. By leveraging the power of separate environments, users can create distinct partitions for different collaborators, control runtime permissions, and establish shared spaces within a single canvas. 

It's worth mentioning that the user experience of collaboration within a computational canvas primarily emphasizes brainstorming and gathering insights, such as generating images, tables, and other outputs. Rather than focusing on co-developing and producing reproducible, reliable code, the platform facilitates the exchange of ideas and knowledge among collaborators.

\section{AI assistance in Computational Canvas}

AI assistants (AI agents) based on foundation models~\cite{shavit2023practices} have demonstrated impressive coding and data analysis abilities. With coding proficiency and advanced reasoning skills, cooperative data exploration and analysis with AI agents could be performed even using natural language only. It allows users to focus more on value than the exact implementation. However, integrating them seamlessly into a user's workflow within notebooks presents a significant challenge in terms of user experience~\cite{weber2024computational}. For example, while debugging in a notebook with agents, users often could experience a cognitive overload due to its fast and autonomous interaction with the notebook~\cite{grotov2024debug}. Thus, it is necessary to ensure that the agent assists users rather than fully controls their workflow. 

Various types of AI agents could be incorporated into the computational canvas, enhancing user productivity while keeping the user in control. These could include \textbf{data analysis assistants} that provide insights and visualizations, \textbf{documentation helpers} that automatically generate explanations for code cells or organize the outputs, \textbf{debugging agents} that analyze errors in the secure environment and suggest fixes, and \textbf{collaborative brainstorming agents} that could offer creative ideas during experimentation. Figure~\ref{fig}(c) shows the example of user interaction with the agent, which could address the requests by creating a separate environment, executing code autonomously, and presenting the results to the user.

Computational canvases provide an interactive platform for cooperative experimentation with autonomous collaborators. By offering a secure, dedicated environment, users can safely grant permissions for code execution and debugging while also facilitating collaborative brainstorming sessions. The two-dimensional nature of the canvas allows for the indication and grouping of cells used by an agent for a particular request in a natural, comprehensible manner. 

\section{Canvas Integration in IDE}
The computational canvas could be a great assistant for developers in IDEs. We've implemented Computational Canvas as a plugin for Visual Studio Code\footnote{https://code.visualstudio.com/}.
To create a canvas within the IDE, users simply need to create a new file with the extension \texttt{.2dntb}. If the file is initially empty, it will be automatically filled with the necessary content upon the first execution, making it ready for use as a canvas file. Additionally, we have ensured that canvas files are compatible with standard Jupyter Notebook \texttt{.ipynb} files. The user can either load the existing notebook into the canvas or save the canvas as a Jupyter notebook with the same cell order as they were created in the canvas. The example of the canvas user interface in the VS Code is shown in Figure~\ref{fig:detached-outputs}.

Much like Project Jupyter, a dedicated server runs whenever a cell is executed within the canvas. The server can manage multiple canvases simultaneously. When a user executes a cell, the client sends a POST request to the server, which then processes the code using a Python REPL environment. When a cell is executed in a newly created environment, the server forks a new Python REPL session from the main one. 

Furthermore, beyond the intuitive user interface, we have introduced the ability to interact with the canvas through an API. This advancement allows AI Agents to interface with the canvas, keeping the agent's logic separate from the canvas. This also could allow other developers to add their own agents easily.

\section{Conclusion}

The evolution of computational notebooks has often led to being overwhelmed for users in the new era of literate programming. The Computational Canvases, a two-dimensional platform, is a promising candidate for the better in-IDE experience of intensive experimentation and data analysis. In this work, we outlined the key features of Computational Canvases, demonstrating how they balance simplicity with functionality, supporting exploration, iterative experimentation, collaborative work, and seamless human-AI interaction in IDE.

By transitioning from linear notebooks to unordered, two-dimensional interfaces, Computational Canvases empower users to focus more on gaining insights and less on working with the code in the common script manner.

We believe Computational Canvases can become an indispensable tool for developers, data scientists, and researchers. Future research could explore the human experience while working on canvases and the patterns while experimenting with the canvas. Ultimately, computational canvases open new frontiers in interacting with code, data, and ideas, opening the way for more creative development experiences. 

\bibliographystyle{unsrt}
\bibliography{bibliography}

\vspace{12pt}

\end{document}